\documentclass[superscriptaddress,prd,aps,preprint,eqsecnum,floats]{revtex4}
\usepackage{graphicx}
\usepackage{epsfig}
\usepackage{amsfonts}
\usepackage{amssymb}
\usepackage{amsmath}
\usepackage{bm}
\usepackage{bbold}
  \newcommand\beq{\begin{equation}}
  \newcommand\eeq{\end{equation}}
  \newcommand\beqn{\begin{eqnarray}}
  \newcommand\eeqn{\end{eqnarray}}

   \newcommand {\nc} {\newcommand}

  \nc {\f} {\frac}                \nc {\s} {\sqrt}
   \nc {\gap} {\\[1ex]}
  \nc {\kap} {\kappa}		% \nc {\kap} {\kappa_p}
  \nc {\amp}[1] {\phi_{#1}}
  \nc {\sgmtot} {\sigma_{\rm tot}}
  \nc {\aml}[2] {\phi_{#1}^{\mathrm{#2}}}
  \nc {\delsigL} {\Delta\sigma_{_{\mathrm L}}}  % for LaTeX; not REVTeX
  \nc {\delsigT} {\Delta\sigma_{_{\mathrm T}}}  % for LaTeX; not REVTeX

\def\pnul{\raise-.3ex\hbox{$\stackrel{\circ}{p}$}}\relax
\def\snul{\raise-.3ex\hbox{$\stackrel{\circ}{s}$}}\relax
\relax

\begin{document}
\title{\begin{flushright}{\small BNL-HET-07/18\\ September 13, 2007 \\ \vspace{1 in} \endflushright}\begin{center} \bf{CNI polarimetry with ${}^3He$}
 \end{center}} 
\normalsize
\author{T. L. Trueman}
\address{Physics Department, Brookhaven National Laboratory, Upton, NY 11973}
\begin{abstract}
By making use of previous analysis of CNI for $pp$ and $pC$ scattering, the spin-flip factor for $np$ scattering is determined as a function of energy and then used to calculate the  $p\,{}^3He$ asymmetry $A_N(s)'$ arising in $p\,{}^3He$ elastic scattering. It is found to be comparable to $A_N(s)$ for $pp$ scattering, but of the opposite sign. It seems that this method could be a practical for measuring the polarization of a ${}^3He$ beam.
\end{abstract}
\maketitle

The Coulomb-nuclear interference enhancement (CNI) at small $-t$ has been studied for high energy proton polarimetry for  some time and it seems interesting to see how it would work for $^3He$. Nigel Buttimore and Elliot Leader and I looked at this a little about five years ago as part of a program to find an absolute polarimeter for protons \cite{BLT}.

Because $^3He$ is also has spin $1/2$ the formalism is similar to $pp$ \cite{2004}; further since the spin of the $^3He$ nucleus is carried by the neutron to a large extent, we will think about this asymmetry as being a polarization measurement of the neutron. Here we will do a very rough calculation which can be improved in several obvious ways.

The most striking difference from $pp$ is that here there are six rather than five amplitudes, the new one corresponding to the neutron ($^3He$) flip and designated as $\phi_6(s,t)$. We can rather generally write (we will neglect $\phi_2, \phi_4$  and set $\phi_3 = \phi_1 =\phi_{+}$ throughout)
\beqn
\phi_{+}^{p\, {}^3He}(s,t) = \frac{3 s}{8 \pi} \sigma_{pp}(s) (i + \rho_{pp}(s)) F_{H}(t) , \nonumber \\
\phi_5^{p\,{}^3He}(s,t) = \frac{ \tau_p  \sqrt{-t}}{m}\phi_{+}^{p\, {}^3He}(s,t),  \nonumber \\
\phi_6^{p\,{}^3He}(s,t) =-\frac{\tau_n  \sqrt{-t}}{3\, m}\phi_{+}^{p\, {} ^3He}(s,t).  \nonumber
\eeqn

The two single spin symmetries, for the proton and the neutron, are
\beqn
A_N \f{d \sigma}{d t}&=&-\f{8 \pi}{s} \mathrm{Im}( \phi_{+}\, \phi_5^*), \nonumber \\
A_N'\f{d \sigma}{d t}&=& \f{8 \pi}{s}  \mathrm{Im}( \phi_{+}\, \phi_6^*). \nonumber
\eeqn
$F_H$ can be calculated by standard  Glauber methods. Here we use simply the harmonic oscillator form $F_H(t) =\exp(t(B/2 +a^2/4))$ with $a^2=57.4 GeV^{-2}$ for illustration \cite{collard}.The $pp$ total cross section $\sigma_{pp}$, the shape parameter $B$ and and the real to imaginary ratio $\rho_{pp}$ are taken from elastic $pp$ data \cite{RPP}. The value of $\tau_p$ is reasonably well measured at  $p_L =24\, GeV/c$ and $p_L=100\,GeV/c$ \cite{pp24}.

\newpage
\samepage{
We still need $\tau_n$. We can get an approximate idea of its size 
from data obtained by the RHIC polarimeter group: the elastic scattering can be thought of as taking place through the exchange of $I=0$  and $I=1$ particles (or Regge poles); it is known that the $I=1$ contribution to the non-flip scattering is very small so we can write (approximately)
\beq
\tau_0 = \tau_p + \tau_n.
\eeq
We can also write
\beq
\tau_{pC}=\tau_0
\eeq
which is obtained by the RHIC polarimeter group's proton-carbon measurements at $p_L =21.7\, GeV/c$ and $p_L=100\,GeV/c$ \cite{pp24} . So we can determine $\tau_n$ at these two energies at least from
\beq
\tau_n=2 \tau_{pC} -\tau_p
\eeq 

Alternatively, we can use a fit I made to both sets of measurements  using \nolinebreak a Regge model which then gives $\tau_n(s)$ as shown in Fig.~1:
\begin{figure}[h!]
\includegraphics{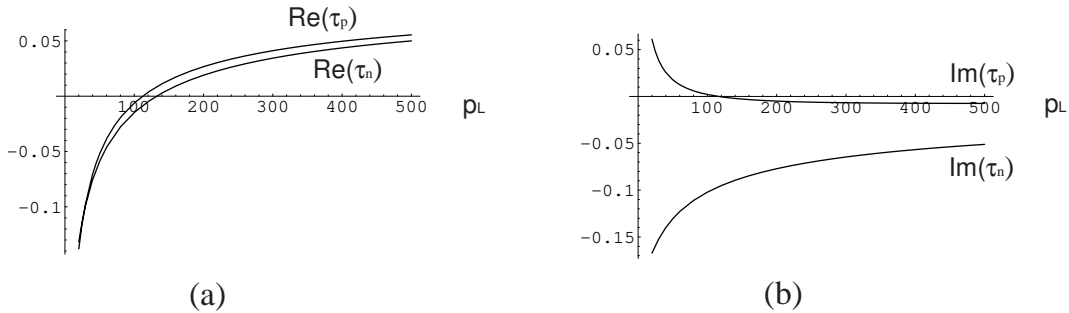}															\caption{Energy Dependence of real and imaginary parts of $pp$ and $np$ spin-flip factors.}
\end{figure}

In order to calculate the CNI analyzing power we need the e.m. amplitudes \cite{BLT}:
\beqn
\phi_+^{em}(s,t) &=&  \frac{2s\, \alpha}{t} \, F_{em}(t), \nonumber
 \\ 
\phi_5^{em}(s,t) &=& -\frac{2s\, \alpha}{ 2m \sqrt{-t}}\, \kappa_p F_{em}(t), \nonumber
 \\
\phi_6^{em}(s,t) &=& \frac{ s\,\alpha}{2 m \sqrt{-t}}\, \kappa_n F_{em}(t) ,  \nonumber
\eeqn
where $F_{em}(t)=\exp{(a^2\,t/4)}$.Ê
Note that $m$ in all these formulas denotes the proton mass. In the second of
these equations one might want to use the ${}^3He$ ion mass along with the magnetic moment of ${}^3He$ rather
than $\kappa_n=-1.91$ \cite{Nigel}. We leave the
expressions this way for consistency with our simple approach.

Let's look first at the two asymmetries in the absence of hadronic spin flip, Fig.~2:

\begin{figure}[h]
\includegraphics{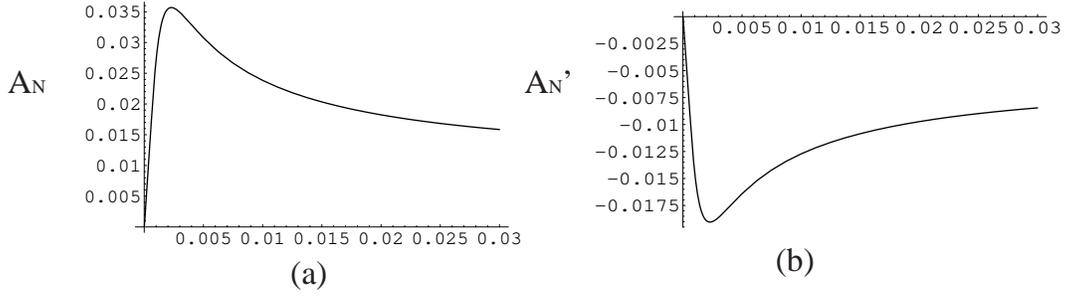}
\vspace{-2.5in}
\caption{Analyzing powers $A_N$ and $A_N'$ at $p_L=100 GeV/c$ with zero hadronic spin-flip factor}	
\end{figure}

Now look at the same things in Fig.~3 using the $\tau$-values found from the $pp$ and $pC$ analysis, Fig.~1:
\begin{figure}[bth]
\includegraphics{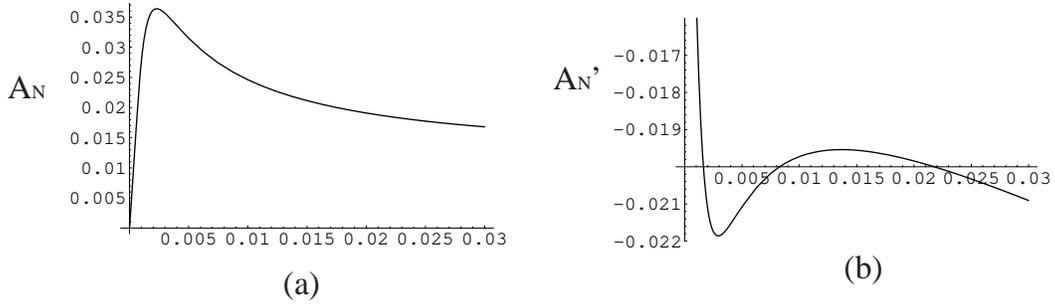}	
\caption{Analyzing powers $A_N$ and $A_N'$ at $p_L=100 \, GeV/c$ using the non-zero hadronic spin-flip factors in text.}	
\end{figure}
we see that the hadronic spin flip significantly modifies the shape of the analyzing power curve especially for the neutron and, very important, both asymmetries are large enough that they should be readily measurable.

From our experience with $pp$ and $pC$ we would expect at least a 10\% measurement of polarization to be possible in this way and should be applicable to colliding beams \cite{finalrep}. An estimate of this for the neutron for 150 GeV colliding beams of $p$ on $^3He$ is shown in Fig.~4.
The method can be extended to ${}^3He-{}^3He$ scattering. 
\begin{figure}[thb]
\includegraphics{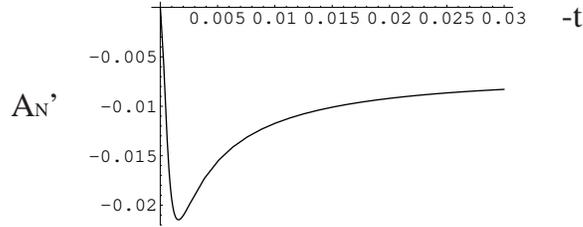}
\caption{Analyzing power $A_N'$ for colliding beams of protons with momentum of $150\, GeV/c$ on a beam of ${}^3He$ at beam momentum $P=3\times150\, GeV/c$. }	
\end{figure}

Thanks to Nigel Buttimore for some important discussions on this topic.


\begin{thebibliography}{99}
\bibitem {BLT} N.~H.~Buttimore, E.~Leader, and T.~L.~Trueman, Phys.Rev. {\bf D64} (2001) 094021
% 
\bibitem{2004} T.~L.~Trueman, Proc. 16th International Spin Physics Symposium  SPIN2004.
%
\bibitem{collard} H.~Collard et al Phys. Rev. 138, {\bf B 57} (1965).
%
\bibitem{RPP} Review of Particle Physics {\bf 592} (2004).
%
\bibitem{pp24}  H. Okada et al., Proc. 17th International Spin Physics Symposium SPIN06, p.681.
arXiv:hep-ex 0704.1031v1
%
\bibitem{Nigel} N.~H.~Buttimore, Proceedings of Spin 2002, BNL,  ed. Y.Makdisi,page 844.
%
\bibitem{finalrep} T.~L.~Trueman "Spin asymmetries for elastic proton scattering and the pomeron spin-dependent couplings",  BNL-HET-07-14 (to be published).
%

\end{thebibliography}
\end{document}